\documentclass[useAMS,usenatbib,referee]{biom}
\usepackage{amsmath}
\usepackage{graphicx,psfrag,epsf}
\usepackage{enumerate}
\usepackage{amsmath,amsfonts,amssymb}
\usepackage{algorithm}
\usepackage{algpseudocode}
\usepackage{dsfont}
\usepackage{url}
\usepackage{setspace}
\usepackage[dvipsnames]{xcolor}
\usepackage{hyperref}
\usepackage{booktabs}
\usepackage{placeins} 

\usepackage{xparse}
\ExplSyntaxOn
\NewDocumentCommand{\longdash}{ O{2} }
 {
  --\prg_replicate:nn { #1 - 1 } { \negthinspace -- }
 }
\ExplSyntaxOff

\usepackage{tabularx}

\newcommand\clearrow{\global\let\rowmac\relax}
\clearrow

\renewcommand{\P}{\mathds{P}} 

\title{Zero-Inflated Bayesian Multi-Study Infinite Non-Negative Matrix Factorization}

\author{Blake Hansen$^{1}$, Dafne Zorzetto$^{2}$, Valeria Edefonti$^{3,4}$, and
Roberta De Vito$^{1,2,5,*}\email{roberta.devito@uniroma1.it}$\\
$^{1}$ Department of Biostatistics, Brown University, Providence, Rhode Island, U.S.A. \\
$^{2}$ Data Science Institute, Brown University, Providence, Rhode Island, U.S.A..\\
$^{3}$ Department of Clinical Sciences and Community Health, Universit\`a degli Studi di Milano, Milano, Italy            \\
$^{4}$ Fondazione IRCCS Ca' Granda Ospedale Maggiore Policlinico, Milano, Italy  \\ 
$^{5}$ Department of Statistical Science, Sapienza University of Rome, Rome, Italy }

\begin{document}









\label{firstpage}


\begin{abstract}
Understanding the association between dietary patterns and health outcomes, such as the cancer risk, is crucial to inform public health guidelines and shaping future dietary interventions. However, dietary intake data present several statistical challenges: they are high-dimensional, often sparse with excess zeros,  and exhibit heterogeneity driven by individual-level covariates. Non-Negative Matrix Factorization (NMF), commonly used to estimate patterns in high-dimensional count data, typically relies on Poisson assumptions and lacks the flexibility to fully address these complexities.  Additionally, integrating data across multiple studies, such as case-control studies on cancer risk, requires models that can share information across sources while preserving study-specific structure.

In this paper, we introduce a novel Bayesian NMF model that (i) jointly models multi-study count data to enable cross-study information sharing, (ii) incorporate a mixture component to account for zero inflation, and (iii) leverage flexible Bayesian non-parametric priors for characterizing the heterogeneity in pattern scores induced by the individual covariates. This structure allows for clustering of individuals based on dietary profiles, enabling downstream association analyses with health outcomes. Through extensive simulation studies, we demonstrate that our model significantly improves estimation accuracy compared to existing Bayesian NMF methods. 
 We further illustrate its utility through an application to multiple case-control studies on diet and upper aero-digestive tract cancers, identifying nutritionally meaningful dietary patterns. An R package implementing our approach is available at \href{https://github.com/blhansen/ZIMultiStudyNMF}{github.com/blhansen/ZIMultiStudyNMF}.

\end{abstract}

%



\maketitle


%

\section{Introduction}
\label{sec:intro}

Diet plays a critical role in the development and prevention of chronic diseases, such as cancer \citep{schulze2018food}. Notably, an estimated 30–40\% of all cancers could be prevented through dietary and lifestyle modifications alone \citep{donaldson2004nutrition}.

Identifying and characterizing dietary patterns is thus critical to understanding the association between diet and disease and to guide public health interventions \citep{edefonti2022diet}. However, dietary behavior is highly heterogeneous across populations and subgroups defined by geography, ethnicity, and cancer types~\citep{stephenson2023racial}. This heterogeneity poses both statistical and substantive challenges, highlighting the need for methods that can simultaneously capture shared dietary structures and subgroups, or study-specific deviations.

Multi-study factor analysis (MSFA) models~\citep{de2019multi, roy2021perturbed, chandra2024inferring,  mauri2025spectral} have emerged as powerful tools for this task offering a principled framework to decompose high-dimensional data into shared and study-specific latent components. These methods have been successfully applied in diverse fields including gene expression analysis \citep{de2021bayesian, grabski2023bayesian}, and more recently nutritional epidemiology \citep{de2019shared, de2025multi}.

However, traditional MSFA methods rely on Gaussian assumptions and are poorly suited for data that are count, sparse, and highly zero-inflated, as is often the case in nutritional epidemiology  (e.g., counts of food items consumed) and in genomics (e.g., somatic mutation counts).

In such settings, standard factor models fail to accurately recover underlying latent patterns, leading to poor model fit, biased factor loadings, and misleading interpretations of latent structure. To address these limitations, Non-negative Matrix Factorization (NMF) has emerged as a powerful alternative for decomposing high-dimensional count data into interpretable latent components~\citep{lee_nmf_paper}. Probabilistic extensions of NMF based on the Poisson distribution~\citep{cemgil_bayesian_2009} have been successfully applied to a wide range of areas such as text mining~\citep{berry_algorithms_2007}, recommender systems~\citep{luo_efficient_2014}, gene expression~\citep{carmona-saez_biclustering_2006, zhang_binary_2007}, cancer genomics~\citep{rosales_signer_2017, islam_uncovering_2022}, and environmental exposure estimation~\citep{rebouillat_estimated_2021}. These models have also been adapted for negative-binomial and ordinal outcomes~\citep{gouvert_nb_nmf_paper}.

Recent work has extended Bayesian NMF to handle multiple studies, enabling the analysis of multi-source count data while retaining the interpretability of non-negative factor models~\citep{grabski2025bayesian}. However, existing NMF approaches, including Bayesian multi-study alternatives, typically fail to account for excess zeros, which are common in many biological and epidemiological datasets.

While zero-inflated NMF models have been proposed~\citep{abe_non-negative_2017}, they are limited to a single-study setting. This gap is particularly limiting in many applications including nutritional epidemiology, where sparsity in consumption data often arises due to non-consumption of certain food items.

Beyond accurately modeling sparse and zero-inflated count data, a second challenge is to associate latent dietary patterns with disease outcomes. This requires accurate estimation of pattern scores, which quantify the degree that each individual expresses on the corresponding patterns, making them particularly relevant in various applications \citep{edefonti2009clustering}. In practice, these scores are often categorized (e.g., quantiles) and used in post-hoc association analyses with disease outcomes~\citep{schulze2018food}.

Flexible and adaptive modeling of pattern scores and clustering of individuals based on this score is essential for accurately estimating association between dietary pattern and disease \citep{morgenstern2021perspective}. Recent work has introduced more flexible priors for pattern scores to account for individual-level heterogeneity \citep{huang2024sparse}, or developed models that directly cluster individuals based on their food consumptions~\citep{fahey2007conditional, stephenson2020robust}. However, existing methods do not incorporate covariate information and do not allow for clustering or adjustment for confounders based on the pattern scores themselves. This limitation is particularly problematic in nutritional epidemiology, where dietary behaviors are strongly influenced by potential confounders such as sex, age, socioeconomic status, body size, and total energy intake \citep{fahey2007conditional}.
Ignoring such factors can obscure true relationships and compromise the interpretability and validity of estimated dietary patterns.

To address these limitations, we introduce a novel Bayesian zero-inflated multi-study NMF (ziNMF) model that extends probabilistic NMF in several important directions: 1- it accommodates zero-inflated count data with study- and variable-specific excess-zero probabilities; 2- it models pattern scores using a nonparametric prior informed by covariates through a dependent probit stick-breaking process \citep{rodriguez2011nonparametric}; and 3- it induces latent clusters of individuals based on pattern scores, thereby enabling interpretable subgroup discovery and improved association with outcomes.  Our method offers the added advantage of simultaneously clustering individuals and learning dietary patterns, providing a more statistically principled and flexible framework than previous approaches in nutritional epidemiology. This enables intuitive model-based association between dietary patterns and cancer.
Specifically, we adopt a Dependent Dirichlet Process prior \citep{quintana2020dependent} for the pattern scores, allowing a flexible, shared prior among all pattern scores given a specific pattern and study. Our prior, informed by covariates via a probit stick breaking process, enables the discovery of latent clusters in the distribution of pattern scores grouping individuals with similar characteristics and dietary behaviors—and facilitates improved estimation of latent confounding structures.

We demonstrate that ziNMF significantly improves estimation, pattern selection, and subject-level interpretation compared to existing Bayesian multi-study NMF approaches, particularly in sparse, heterogeneous, and zero-inflated settings. Finally, we apply ziNMF to a real-world multi-study dataset on dietary intake and cancer outcomes, collected from three large Italian case-control studies on upper aero-digestive tract cancers (oral cavity and pharynx, larynx, and esophagus), highlighting its ability to identify meaningful dietary patterns and simultaneous clusters of individuals, validated with a case/control cancer incidence rate. 

The plan of the paper is as follows. Section~\ref{sec:model} defines the statistical model and priors used to define Zero-Inflated Bayesian Multi-Study NMF, Section~\ref{sec:simulations} provides comprehensive simulations benchmarking our method against state-of-the-art Bayesian NMF algorithms, Section~\ref{sec:application} applies ziNMF to dietary data collected from three Italian case-control studies, and Section~\ref{sec:Discussion} discusses the implications of our findings and outlines directions for future research.

\section{Zero-Inflated Infinite Multi-Study Non-Negative Matrix Factorization}
\label{sec:model}

Let us consider observed count data $\mathbf{M}_{s} \in \mathcal{Z}_+^{P \times N_s}$, for each study $s\in \{1,\dots,S\}$, where study $s$ consists of $N_s$ observations measured on the same set of $P$ variables. Additionally, for each study $s$, we observe $Q_s$ covariates collected in the matrix $\mathbf{X}_s \in \mathbb{R}^{N_s \times Q_s}$.  For each individual $i =1, \dots, N_s$ we denote the covariate vector as $\mathbf{x}_{is}=(1, x_{is2}, \dots, x_{isQ_s})^{\intercal}$, including the intercept for notational simplicity.

 To model this correlated, high-dimensional count data in $\mathbf{M}_s$, we adopt a probabilistic NMF framework based on the Poisson distribution \citep{cemgil_bayesian_2009}.
However, since the data exhibit substantial zero inflation, we extend this framework by incorporating a zero-inflated Poisson component via a mixture distribution. Specifically, for each variable $p$, individual $i$, and study $s$, we model:
\begin{align}
    M_{pis} &= \pi_{ps} \delta_0  + \left(1-\pi_{ps}\right)\text{Poisson}\left(\left[\mathbf{W}\mathbf{H}_s\right]_{pi} 
    \right),
    \label{eq:def_model}
\end{align}
where $\mathbf{W} \in \mathcal{R}_+^{P\times K}$ is the pattern loading matrix, and $\mathbf{H}_s$ is the latent score matrix. Each element $h_{kis}\in\mathcal{R}_+$ quantifies the contribution of the latent pattern $k$ to the observed counts vector for the individual $i$ of study $s$. 

Specifically, we can rewrite the matrix decomposition in \eqref{eq:def_model} as
\begin{equation*}
    \left[\mathbf{W}\mathbf{H}_s\right]_{pi} =\sum_{k=1}^K w_{pk}h_{kis},
\end{equation*}
where we ensure that $\{w_{pk}\}_{k=1}^K\in[0,1]^K$ and $\sum_{k=1}^K w_{pk}=1$ for each variable $p$.

The zero-inflation probability $\pi_{ps}$ is independent of the factorization and have the following prior:
\begin{equation*}
    \pi_{ps} \sim \text{Beta}\left(\alpha^m, \beta^m\right). 
\end{equation*}

For the pattern matrix $\mathbf{W}$, we assume independent Gamma priors~\citep{cemgil_bayesian_2009, rosales_signer_2017, grabski2025bayesian}:
\begin{equation*}
    w_{pk} \sim \text{Gamma}\left(\alpha^w, \beta^w\right).
\end{equation*}

To improve flexibility in modeling covariate dependent heterogeneity in the score matrix $\mathbf{H}s$, we adopt a Bayesian nonparametric (BNP) framework via a covariate-dependent Dirichlet process prior on each score.
Specifically, we leverage the Bayesian non-parametric (BNP) approach and adopt a dependent Dirichlet process \citep[DDP; ][]{mac2000dependent, quintana2020dependent} as prior for each $h_{ski}$:
\begin{gather}
    h_{kis}\mid \mathbf{x}_{is} \sim f_{kis}\left(\cdot \mid \mathbf{x}_{is}\right),\notag \\
     f_{kis}\left(\cdot \mid \mathbf{x}_{is}\right) = \int_{\Psi}\mathcal{K}\left(\cdot\mid \psi \right) d G_{k, \mathbf{x}}\left(\psi\right),\label{eq:score_BNPprior} \\
     G_{k, \mathbf{x}}\left(\psi\right) \sim \Pi_{k, \mathbf{x}}, \notag
\end{gather}
where $\mathcal{K}\left(\cdot\mid \psi \right)$ is a kernel density with parameter $\psi \in \Psi$ and $G_{k, \mathbf{x}}$ is a random probability measure specific to pattern $k$ and dependent on the observed covariate vector $\mathbf{x}$. We assume $\Pi_{k, \mathbf{x}}$, a pattern-specific and covariate dependent nonparametric process, as prior for $G_{k, \mathbf{x}}$.

Using the single-atom representation \citep{quintana2020dependent}, we can write the random probability measure as an infinite mixture, as follows:
\begin{align*}
    G_{k, \mathbf{x}} &= \sum_{l=1}^\infty \omega_{kl}\left(\mathbf{x}_{is}\right) \delta_{\psi_{kl}},
\end{align*}
with $\{\omega_{kl}(\mathbf{x}_{is})\}_{l\geq 1}$ denoting the infinite sequences of random weights for each pattern $k$ and dependent on the covariate $\mathbf{x}_{is}$, and $\{\delta_{\psi_{kl}}\}_{l\geq 1}$ are the infinite sequences of random kernel’s parameter. 

Equation~\eqref{eq:score_BNPprior} can be re-written as:
\begin{align*}
    h_{kis}\mid \mathbf{x}_{is}, \psi &\sim \sum_{l=1}^\infty \omega_{kl}\left(\mathbf{x}_{is}\right)\mathcal{K}\left(\cdot\mid \psi_{kl} \right).
\end{align*}

For the infinite sequences of random weights, we adapt the dependent probit stick-breaking process, introduced by \citet{rodriguez2011nonparametric}, to each latent pattern $k$, such that:
\begin{gather*}
    \omega_{kl}\left(\mathbf{x}_{is}\right)= \Phi\left(\alpha_{kl}(\mathbf{x}_{is})\right)\prod_{r<l}\left\{\Phi\left(1-\alpha_{kr}(\mathbf{x}_{is})\right)\right\}, \\
    \alpha_{kl}(\mathbf{x}_{is}) \sim \mathcal{N}\left(\mathbf{x}_{is}^\intercal \boldsymbol{\beta}_{kls},1\right), 
\end{gather*}
where, we assume $\boldsymbol{\beta}_{kls} \sim \mathcal{N}\left(\boldsymbol{\beta}_{l}^{0}, \tau^{0}_{l}\text{I}_{Q_s}\right)$, with $\text{I}_{Q_s}$ indicating the identity matrix with dimension $Q_s$.

We assume a Gamma distribution for the random kernel $\mathcal{K}\left(\cdot\mid \psi_{kl} \right)$, with $\psi_{kl}=(c_{kl}, \theta_{kl})\in\mathcal{R}_+^2$, yielding: 
\begin{gather}
    \mathcal{K}\left(h_{kis} \mid \psi_{kl} \right) = \mbox{Gamma}(c_{kl},c_{kl}/\theta_{kl}),
\end{gather}
where the mean is equal to $\theta_{kl}$ and the variance is $\theta_{kl}/c_{kl}$.

The Gamma distribution is a natural and widely adopted choice in the NMF literature \citep{cemgil_bayesian_2009}, as it ensures the non-negative real domain for the scores $h_{kis}$.
Moreover, our use of the Gamma kernel enables the model  to identify scores close to zero, which serves two key purposes: (i) it induces shrinkage toward zero, promoting sparsity in the latent representations, and (ii) it facilitates the automatic identification of study-specific significant patterns in each study $s$.
  To achieve these goals, we constrain the parameters of the first component of the mixture as follows:
\begin{gather*}
    c_{k1}=1 \quad \mbox{and} \quad \theta_{k1}=\epsilon<0.5,
\end{gather*}
while the following component of the mixture has parameters equal to
\begin{gather*}
    c_{kl}=c\geq2 \quad \mbox{and} \quad \theta_{kl} \in \mathcal{R}_+ \quad \forall l>2,
\end{gather*}
where $c$ is a fixed hyperparameter. We assume the following prior distributions:
\begin{equation*}
    \theta_{kl}\sim \mathcal{IG}\left(\gamma_1^\theta, \gamma_2^\theta\right),
\end{equation*}
where $\mathcal{IG}$ denotes the inverse gamma distribution with $(\gamma_1^\theta, \gamma_2^\theta) \in \mathcal{R}_+^2$, mean $\gamma_2^\theta/(\gamma_1^\theta-1)$, and variance $(\gamma_2^\theta)^2/[(\gamma_1^\theta-1)(\gamma_1^\theta-2)]$.

The full model and its posterior distributions define the Zero-Inflated Multi-Study NMF model, which we call 'ziNMF' for the remainder of this manuscript.


\subsection{Pattern selection}
\label{subsec:selection}

In this section, we show how the specification of a mixture distribution as a prior for the pattern scores, particularly through parameter constraints on the first mixture component, enables automatic identification of relevant patterns within each study. Before proceeding, we introduce some notation to facilitate the explanation. 

Due to the discrete nature of the random probability measure $G_{k, \mathbf{x}}$ introduced in Section~\ref{sec:model}, each score $h_{kis}$ can be associated with a latent cluster indicator.  Specifically, for each pattern $k$, individual $i$, and study $s$, we define a latent multinomial variable $\mathbf{D}_{kis}=(D_{k1is}, \dots, D_{kL^*is})$ of dimension $L^*$, where $D_{klis}=1$ if $h_{kis}$ belongs to the cluster $l$, and $0$ otherwise. This implies:
\begin{equation*}
    \Pr(D_{klis}=1 ) =  \omega_{kl}\left(\mathbf{x}_{is}\right).
\end{equation*}

The $L^*< \infty$ indicates the upper bound for the number of components in the mixture. This truncation, if chosen large enough, preserves the properties of the processes, as shown by \citet{rodriguez2011nonparametric}, and enables efficient computation for model estimation. 
Conditional on the cluster allocation $\mathbf{d}_{kis}$, we can express the distribution of $h_{kis}$ from Eq.~\eqref{eq:score_BNPprior} as:
\begin{equation*}
   \{ h_{kis} \mid d_{klis}=1, \mathbf{d}^{(-l)}_{kis}=0, \psi\}\sim \mbox{Gamma}(c_{kl},c_{kl}/\theta_{kl}),
\end{equation*}
where the unit $i$ in the study $s$ has been allocated in the cluster $l$ for the pattern $k$. The notation $\mathbf{d}^{(-l)}_{kis}$ denotes the vector $\mathbf{D}_{kis}$ with the  $l$-th element removed, and $h_{kis}$ is now conditionally Gamma-distributed given its cluster assignment.

Therefore, the probability to remove the pattern $k$ in the study $s$ is proportional to the probability of belonging to the first component of the mixture and characterized by the observed covariates $\mathbf{x}_{is}$:
\begin{gather}
    \Pr(\mathbf{D}_{k1\cdot s}= 1) =\prod_{i=1}^{n_s} \omega_{k1}\left(\mathbf{x}_{is}\right) = \prod_{i=1}^{n_s} \Phi\left(\alpha_{k1}(\mathbf{x}_{is})\right).
\end{gather}

\subsection{Posterior Computation}

To perform inference under the proposed model, we develop a fast and efficient Gibbs sampler for posterior inference, publicly available at: \href{https://github.com/blhansen/ZIMultiStudyNMF}{\texttt{github.com/blhansen/ZIMultiStudyNMF}}.

For the purposes of sampling, we augment the model in Eq.~\eqref{eq:def_model} with latent counts $z_{pkis}$ such that $\sum_{k}z_{pkis}=m_{pis}$ and a binary indicator $a_{pis}$ to distinguish between zero-inflated and Poisson components. 
Thus, the  model becomes:
\begin{align}
   z_{pkis} \mid a_{pis} &\sim a_{pis}\delta_0 + (1-a_{pis}) \text{Poisson}\left(w_{pk}h_{kis}\right),
\end{align}
where the binary indicator $a_{pis}$ has probability distribution and priors for the parameters, respectively, equal to:
\begin{gather*}
     a_{pis} \mid \pi_{ps} \sim \text{Bernoulli}(\pi_{ps}), \quad
     \pi_{ps} \sim \text{Beta}(\alpha^m, \beta^m),
\end{gather*}
where $\alpha^m, \beta^m \in \mathcal{R}^2_+$ are fixed hyper-parameters.

\vspace{0.5em}
\textit{Latent Counts.}
By construction, when a unit $i$ in the study $s$ has $a_{spi}=1$ indicating that it belongs to the spike in \eqref{eq:def_model}, the corresponding latent counts $K$ are zeros, such that $z_{p1is}= \dots=z_{pKis}=0$. Otherwise, when $a_{spi}=0$, they are independently Poisson distributed with known sum $\sum_{k} z_{pkis} = m_{pis}$, such that:
\begin{align*}
    \{z_{pkis}\}_{k=1}^K \mid a_{spi}=0,  \dots &\sim \text{Multinomial}\left(m_{pis}, \boldsymbol{\pi}^z_{pis}\right), \\
    \pi^z_{pkis} &\propto w_{pk}h_{kis},
\end{align*}
which allows us to sample $z_{p1is}, \dots, z_{pKis}$ conditional on indicators $\mathbf{A}_s$, pattern matrix $\mathbf{W}$ and scores matrix $\mathbf{H}_s$.

\vspace{0.5em}
\textit{Excess Zero Indicators.}
The posterior probabilities for the indicator variable $a_{pis}$ is a Bernoulli distribution as follows:
\begin{gather*}
a_{spi} \mid \dots \sim \text{Be}\left(\frac{P(a_{spi}=1\mid\dots)}{P(a_{spi}=1\mid\dots) + P(a_{spi}=0\mid\dots)}\right),\\
    \Pr(a_{spi}=1\mid\dots) \propto 
     \pi_{ps}\mathbb{I}_{\left\{m_{pis}=0\right\}} , \\
    \Pr(a_{spi}=0\mid\dots) \propto 
     (1-\pi_{ps}) \text{Poisson}\left(m_{pis}, \sum_{k=1}^K w_{pk}h_{kis}\right),
\end{gather*}
where the excess zero probability parameter $\pi_{ps}$ has a posterior distribution:
\begin{align*}
    \pi_{ps} \mid \dots &\sim \text{Beta}\left(\alpha^m + \sum_{i=1}^{N_s}a_{pis}, \beta^m +N_s - \sum_{i=1}^{N_s}a_{pis}\right).
\end{align*}

\vspace{0.5em}
\textit{Pattern Matrix.} Considering now the elements involved in the Poisson distribution, the posterior distribution for each loading $p$ of pattern $k$ is given by:
\begin{align*}
    w_{pk} &\sim \text{Gamma}\left(\alpha^w  +\sum_{s=1}^S\sum_{i=1}^{n_s} z_{pkis}, \beta^w +  \sum_{s=1}^S\sum_{i=1}^{n_s} h_{ski}(1-a_{pis}) \right).
\end{align*}

\vspace{0.5em}
\textit{Scores Matrix.}
Following \cite{rodriguez2011nonparametric} and as already introduced in Section \ref{subsec:selection}, we truncate the infinite mixture to $L^*$ and introduce the augmentation data, through the latent binary vector $\mathbf{d}_{kis}$ of dimension $L^*$, where $d_{klis}=1$ if $h_{kis}$ belongs to the cluster $l$ and $0$ otherwise. Therefore, we can rewrite the prior as the following:
\begin{equation*}
    p(h_{kis}\mid \mathbf{d}_{kis}, \mathbf{c} , \boldsymbol{\theta}_{k}) = \prod_{l=1}^{L^*} \text{Gamma}\left(c_l, c_l/\theta_{kl}\right)^{d_{klis}},
\end{equation*}
which implies the posterior distribution equal to:
\begin{gather*}
    h_{kis} \mid  \dots \sim \text{Gamma}\left(c_{kis} + \tilde{z}, \frac{c_{kis}}{\theta_{kis}} + \tilde{w}\right),\\
    \tilde{z}=\sum_{p=1}^P z_{pkis}(1-a_{pis}),\quad
    \tilde{\omega} = \sum_{p=1}^P w_{pk}(1-a_{pis}).
\end{gather*}
where $c_{kis}= \mathbf{c}^\intercal\mathbf{d}_{kis}$  and $\theta_{kis}=\boldsymbol{\theta}_{k}^\intercal\mathbf{d}_{kis}$ are equal to $c_l$ and $\theta_{kl}$ for the cluster corresponding to $h_{kis}$.

\vspace{0.5em}
\textit{Cluster Indicators.}
The posterior distribution of the cluster indicator $\mathbf{d}_{kis}$ follows a multinomial distribution with probabilities:
\begin{align*}
    P(d_{klis}=1) &\propto \omega_{klis}\text{Gamma}\left(h_{kis}\mid c_l, \theta_{}\right).
\end{align*}

\vspace{0.5em}
\textit{Cluster-Specific Mean.}
The posterior distribution for the Gamma parameter $\theta_{kl}$, indicating the cluster-specific mean, for $l=2,\dots,L^*$ is given by:
\begin{align*}
    \theta_{kl}\mid \dots &\sim \mathcal{IG}\left(\gamma_1^\theta + c_l\sum_{s=1}^S\sum_{i=1}^{N_s}d_{klis}, \gamma_2^\theta + c_l\sum_{s=1}^S\sum_{i=1}^{N_s} h_{kis}d_{klis}\right).
\end{align*}

\vspace{0.5em}
\textit{Probit Regression Parameters.}
Following \citet{albert2001sequential}, to obtain a conjugate posterior for the covariate regression parameters $\beta_{kls}$ in the probit, we need to introduce the augmented variables $y_{klis}$ with full conditional distribution for $l=1,\dots,L^*-1$:
\begin{align*}
    y_{klis} \begin{cases}
        \mathcal{N}_+\left(\mathbf{x}_{is}^\intercal \boldsymbol{\beta}_{kls},1\right) & d_{klis} = 1, \\
         \mathcal{N}_-\left(\mathbf{x}_{is}^\intercal \boldsymbol{\beta}_{kls},1\right) & \sum_{r=1}^{l} d_{kris} = 0,
    \end{cases}
\end{align*}
where $\mathcal{N}_+(\mu, 1)$, $\mathcal{N}_{-}(\mu, 1)$ denotes the normal distribution truncated above or below zero, respectively, allowing us to sample positive values for $y_{klis}$ when $h_{kis}$ belonging to cluster $l$ and sample negative values for $y_{klis}$ when $h_{kis}$ belonging to clusters $l+1,\dots, L^*$.

Consequently, the posterior distribution for $\beta_{kls}$ for $l=1,\dots,L^*-1$ is given by:
\begin{gather*}
    \beta_{kls} \sim \mathcal{N}\left(\boldsymbol{\mu}_{kls}, \boldsymbol{\Sigma}_{kls}\right), \\
    \boldsymbol{\Sigma}_{kls} = \left(\mathcal{\tau^0I_{Q_s}} + \tilde{\mathbf{x}}_{s}^\intercal\tilde{\mathbf{x}}_{s}\right)^{-1}, \quad
    \boldsymbol{\mu}_{kls} = \boldsymbol{\Sigma}_{kls}\left(\tau^0I_{Q_s}\boldsymbol{\beta}^0_l + \tilde{\mathbf{x}}_{s}^\intercal\tilde{\mathbf{y}}_{s}\right),
\end{gather*}
where $\tilde{\mathbf{x}}_{s}$, $\tilde{\mathbf{y}}_{s}$ are constructed from $\mathbf{x}_{s}$, $\mathbf{y}_{s}$ to contain individuals such that the corresponding score $h_{ski}$ belongs to cluster $l$ or higher, i.e.:  $\tilde{\mathbf{x}}_{s}=\left\{\mathbf{x}_{is} : \sum_{r=l}^{L^*} d_{kris}=1  \right\}$, $\tilde{\mathbf{y}}_{s}=\left\{\mathbf{y}_{is} : \sum_{r=l}^{L^*} d_{kris}=1  \right\}$.

\section{Simulation Study}
\label{sec:simulations}

To evaluate the utility of our proposed model, we conduct extensive simulation studies assessing its ability  to accurately estimate: i) pattern loadings $\mathbf{W}$, ii) study-specific pattern scores $\mathbf{H}_s$, iii) NMF reconstruction $\mathbf{W}\mathbf{H}_s$, iv) assignment of patterns across studies and covariates, and v) the number of latent patterns.

We consider three simulation scenarios, each with 50 replicates. In Scenario 1, we simulate counts vectors from 5 non-overlapping latent patterns across $S=3$ studies, each with $N_s=100$ observations and $P=20$ variables, with approximately $25\%$ zeros per column. 
Pattern scores $h_{kis}$ are generated from a Gamma distribution with shape 10 and rate 0.2 in approximately $50\%$ of samples in each study according to the study-sharing pattern shown in Figure~\ref{fig:sims_sharing}. 

Scenario 2 maintains the same number of studies and samples but increases the number of variables $P=50$. Here, latent patterns are drawn from a Dirichlet distribution with a constant concentration parameter of $0.5$. Additionally, we introduce 2 covariates: a binary covariate $x_1 \sim \text{Bernoulli}(0.5)$ , and a continuous covariate, $x_2 \sim \mathcal{N}(0,1)$. The scores $h_{kis}$ are drawn from a Gamma distribution with shape 10 and rate 0.2 with probabilities dependent on covariate values. 

In Scenario 3, we set the dimension to mimic a realistic real data application ($S=3$, $P=22$, $N=946, 460, 304$), with the proportion of zeros in each variable to mirror the real data application. Pattern scores are drawn from 3 different clusters according to a Gamma distribution with shape and rate parameters of $(1, 0.25)$, $(10,2)$, and $(10,0.2)$, respectively, where the probability of belonging to each cluster is proportional to the covariate-dependent probit stick breaking process with covariate effects $\boldsymbol{\beta}_{kls} \sim \mathcal{N}(0, \mathcal{I}_{Q_s})$.

We compare our ziNMF model to two Bayesian NMF-based models: 1- Compressive NMF~\citep{zito2024compressive} (indicated with CompNMF in the results), a single-study approach that uses compressive hyperpriors for pattern selection; and 
 2- Multi-Study NMF (indicated with msNMF)~\citep{grabski2025bayesian}, a multi-study approach that can select relevant patterns at the study level. We utilize the Gibbs and Metropolis-within-Gibbs implementations for each model, respectively, publicly available at \href{https://github.com/alessandrozito/CompressiveNMF}{\texttt{github.com/alessandrozito/CompressiveNMF}} and  \href{https://github.com/igrabski/MultiStudyNMF}{\texttt{github.com/igrabski/MultiStudyNMF}}, using the recommended settings. To accommodate multi-study data with CompNMF, we stack the simulated datasets into one large dataset and the analysis proceeds as if they were all observed in the same study. Note that neither benchmark model accounts for zero inflation.

In the following results, we indicate our proposed zero inflated multi-study NMF with ziNMF, estimated with the publicly available code at \href{https://github.com/blhansen/ZIMultiStudyNMF}{\texttt{github.com/blhansen/ZIMultiStudyNMF}}.

\subsection{Estimation Accuracy}

\begin{figure}[h!]
    \centering
    \includegraphics[width=0.8\linewidth]{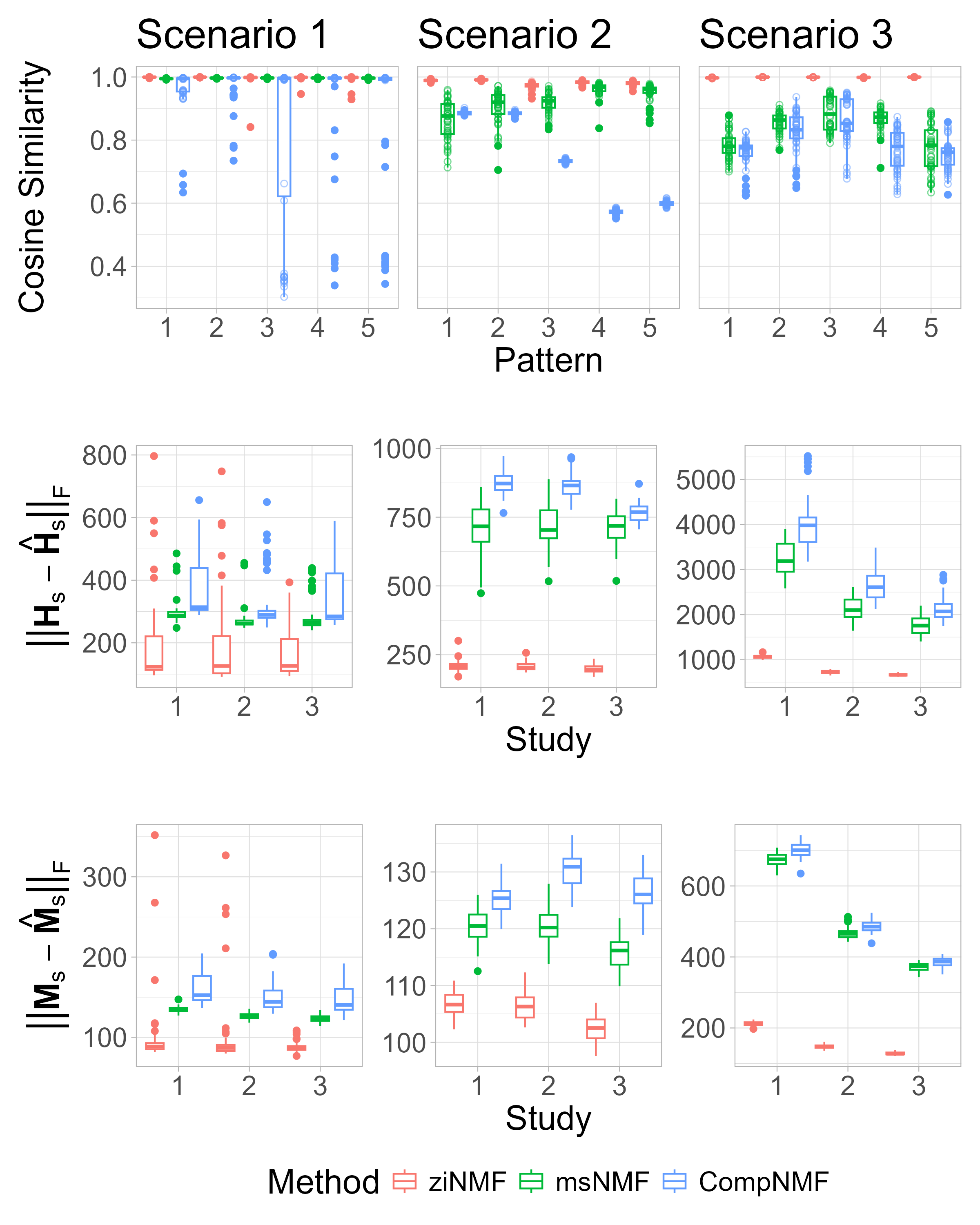}
    \caption{Estimation accuracy across each of 3 simulation scenarios, measured by 1)maximum cosine similarity between true and estimated columns of $\mathbf{W}$, 2) Frobenius norm of the difference between the true and estimated scores $\mathbf{H}_s$, and 3) Frobenius norm of the difference between the true and estimated $\mathbf{W}\mathbf{H}_s$.}
    \label{fig:estimation_accuracy}
\end{figure}

We first assess the accuracy of pattern estimation using cosine similarity: $\max_{k,k'} d_{cos}(\mathbf{W}_k, \hat{\mathbf{W}}_{k'})$, where $d_{cos}(\mathbf{x},\mathbf{y})$ denotes the cosine similarity between two vectors $\mathbf{x}$, $\mathbf{y}$. The results for each simulation scenario are shown in Figure~\ref{fig:estimation_accuracy}. In Scenario 1, both ziNMF and msNMF recover patterns with very high accuracy (cosine similarity $> 0.99$ for most patterns), whereas CompNMF exhibits wide variability, with some simulations dropping as low as 0.4, indicating poor performance. In Scenario 2, where the number of variables increases while the same sample is unvaried, ziNMF maintains high cosine similarities for each of the patterns, while msNMF had a noticeable drop in performance, especially for patterns 1 and 2 (mean cosine similarities of $0.86$ and $0.90$), which were present in the majority of observations for every study. CompNMF had fair estimation for patterns 1 and 2 (average cosine similarity of 0.88 for both patterns), but estimates the study-specific patterns 3-5 poorly (average cosine similarity 0.73, 0.57, 0.60).  Scenario 3 again demonstrates ziNMF's superior recovery (mean $\geq$ 0.99), with msNMF and CompNMF trailing at 0.75--0.88.

Next, we assess the ability of each method to accurately estimate pattern scores, $\mathbf{H}_s$, as well as the reconstruction error of the entire matrix decomposition, $\mathbf{W}\mathbf{H}_s$. We note that in this context, the matrix decomposition $\mathbf{W}\mathbf{H}_s$ represents the mean of the data, $\mathbf{M}_s$, if there are no structural zeros present - meaning approximately $5\%-80\%$ of the values are censored in the simulated data, depending on the simulation scenario and variable. For both metrics, we evaluate each method based on the Frobenius norm of the difference between the estimated and ground truth parameters, as displayed in Figure~\ref{fig:estimation_accuracy}. For the purposes of this experiment, the estimated $\mathbf{H}_s$ are ordered by matching the estimated $\mathbf{W}$ with the ground truth, assigning each estimated pattern to its closest ground truth pattern with a cosine similarity above $0.8$. Since CompNMF assumes a Dirichlet prior on the columns of $\mathbf{W}$, we scale the ground truth and estimate scores to correspond to a normalized estimated pattern matrix, $\tilde{\mathbf{H}}_s$, before calculating the Frobenius norm. 

In each simulation scenario, ziNMF had the lowest error in estimating both $\tilde{\mathbf{H}}_s$ and $\mathbf{W}\mathbf{H}_s$. In Scenario 1, the score matrix for the ziNMF ranges between $46\%-72\%$ of the errors for msNMF and CompNMF on average.  In the same scenario, the reconstruction error for the ziNMF is on average $60\%-80\%$ of the competing methods. In Scenario 2, ziNMF performs even better in terms of estimating the pattern scores, getting errors that are on average only $23\%-29\%$ of the errors of the competing methods. The reconstruction error is slightly more favorable to msNMF and CompNMF, with both methods having reconstruction errors that are approximately $10\%$ higher than ziNMF. This is because in this scenario, there are more than double the measured variables in Scenario 1 ($50$ vs. $20$), but the level of sparsity is still the same ($25\%$ excess zeros in each column of $\mathbf{M}_s$), allowing the methods to estimate the decomposition $\mathbf{W}\mathbf{H}_s$ with greater accuracy. In Scenario 3 ziNMF outperforms all the other competitors, achieving relative errors that are $27\%-37\%$ for scores, and $30\%-34\%$ for reconstruction compared to msNMF and CompNMF, respectively. 

\subsection{Pattern Assignment}

\begin{figure}[h!]
    \centering
    \includegraphics[width=\linewidth]{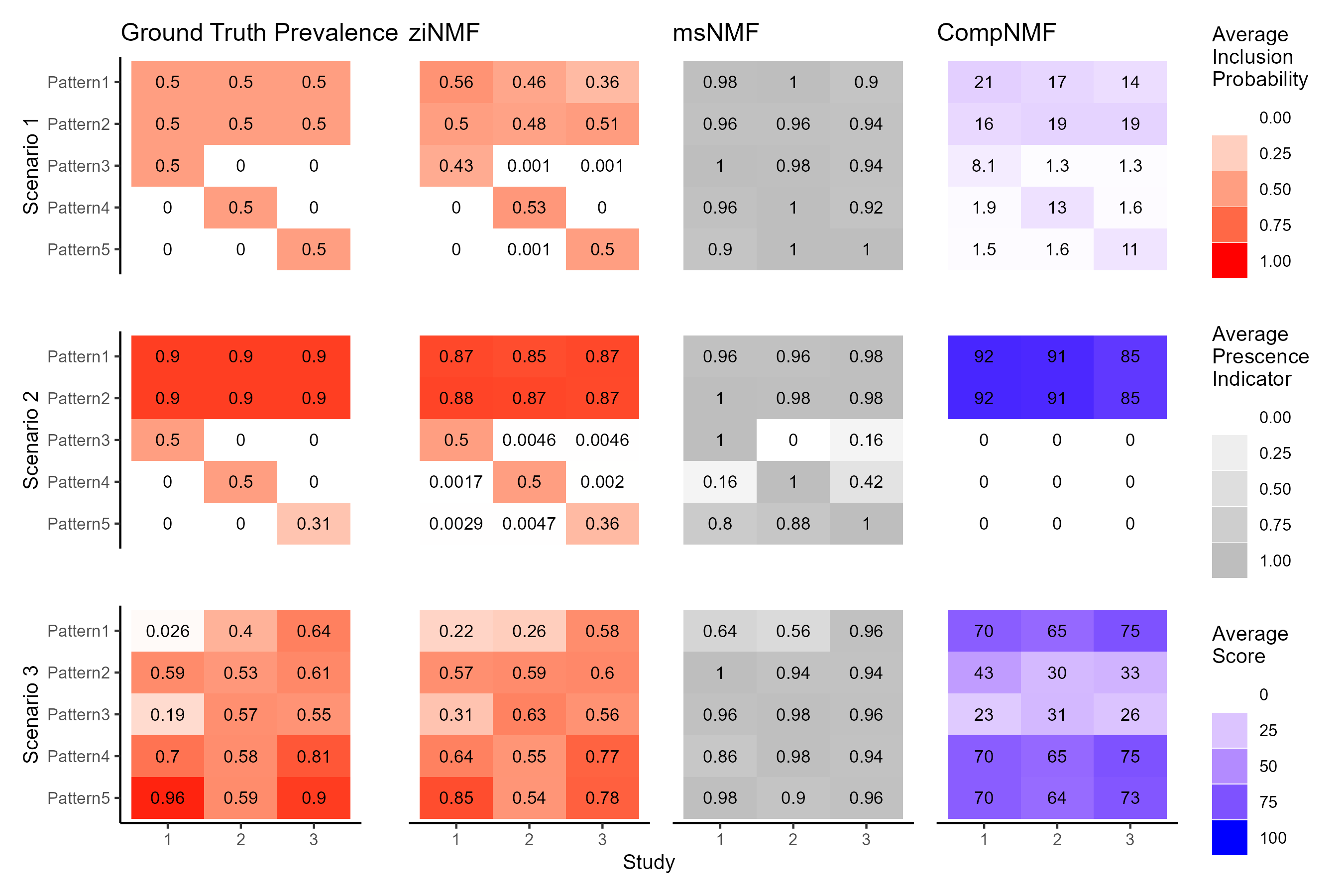}
    \caption{Estimated pattern sharing across $S=3$ studies in each of the three simulation scenarios. Each plots-line corresponds to a simulation scenario.}
    \label{fig:sims_sharing}
\end{figure}

In this section, we assess the ability of each method to correctly assign estimated patterns across studies.  The results in Figure~\ref{fig:sims_sharing} shows the average per-study pattern prevalence. For ziNMF, we report the average posterior probability that individuals are not assigned to cluster 1 of the DPP for pattern scores (recall the specification of the DDP in Section~\ref{subsec:selection}). For msNMF, which uses study-level indicators, $A_{sk}$ denotes if a pattern is present for at least one sample in study $s$, we report the average of this indicator across the $50$ replicates. For CompNMF, which is a single-study method, we report the average pattern score in each study. 

ziNMF consistently captures the true pattern-sharing structure, assigning low probabilities to absent patterns. CompNMF performs well in Scenarios 1 and 3 but fails entirely in Scenario 2, missing all study-specific patterns. msNMF tends to overestimate pattern prevalence by assigning every pattern to every study in most replicates. These results highlight ziNMF's superior flexibility in identifying study-specific and shared patterns.

\subsection{Selecting the number of patterns}

\begin{figure}
    \centering
    \includegraphics[width=0.5\linewidth]{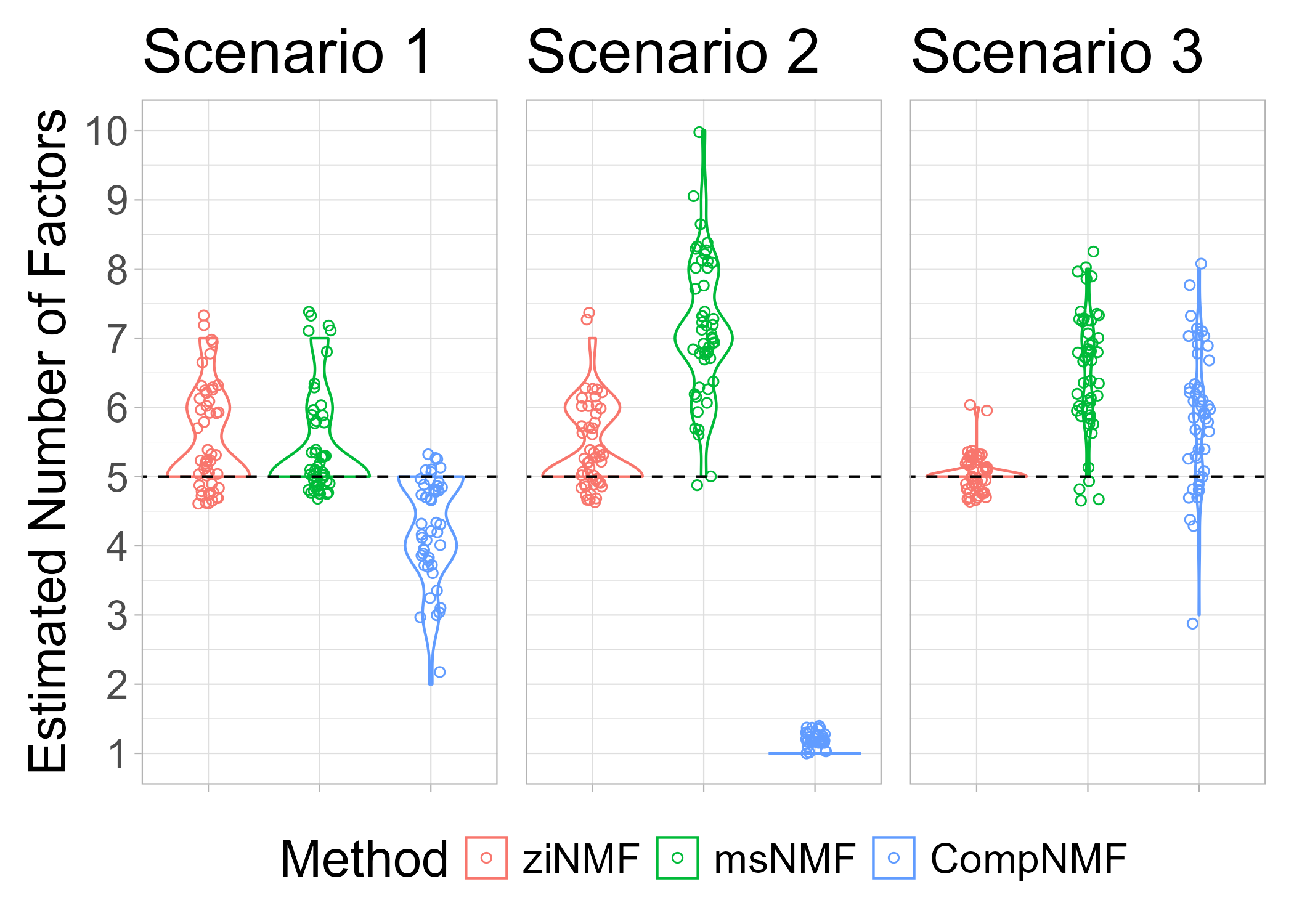}
    \caption{Estimated number of components across replicates for each simulation scenario.}
    \label{fig:sims_n_factors}
\end{figure}

We assess the ability of each method to correctly learn the number of components. msNMF and CompNMF both take a similar approach to ziNMF, beginning sampling with a conservative upper bound on the number of patterns, and then dropping patterns that meet certain criteria during sampling. To make the results as comparable as possible, all models begin with an upper bound of $10$ patterns (true value = 5) . The number of patterns learned in each replicate is reported in Figure~\ref{fig:sims_n_factors}. 

In Scenario 1, both ziNMF and msNMF typically identify between 5 and 7 patterns; CompNMF varies from 2 to 5 patterns. In Scenario 2, ziNMF selects 5-6 patterns, while msNMF overestimates the number of patterns to be mostly between 6-8. In this scenario, the compressive effect of CompNMF is very influential, causing the model to learn 1 pattern for each replicate, which appears to be a combination of patterns 1 and 2. In Scenario 3, ziNMF correctly recovers the true number of patterns in 48 out of 50 replicates, while msNMF tends to estimate 7 or 8 components, and CompNMF estimates between 5 and 7 components. These results indicate that, while selecting the number of components remains a challenging task for NMF-based methods, ziNMF outperforms all the other approaches.

\FloatBarrier
\section{Cancer Case Study Application}

\label{sec:application}
In this section, we apply our ziNMF to dietary data collected from $S=3$ parallel case-control studies investigating the association between dietary habits and upper aerodigestive tract cancers in Italy, as described in detail in \citet{accardi_dietary_2019}. 
These multicentric studies included incident, histologically confirmed cancer cases and corresponding hospital-based controls recruited from the same network of hospitals.   Controls were admitted for a wide spectrum of acute, non-neoplastic conditions unrelated to smoking, alcohol consumption, or long-term dietary modifications. Matching was performed on age (5-year categories), sex, and recruitment center; to compensate for the rarity of laryngeal and esophageal cancers in women, a control-to-case ratio of about 5 was chosen for women and of about 2 for men. All studies followed a common protocol, including inclusion and exclusion criteria, and a standardized interviewer-administered questionnaire on sociodemographic characteristics, anthropometric measures, lifestyle habits, and medical history \citep{accardi_dietary_2019}. 
Dietary intake was assessed via a reproducible and valid $78$-item food-frequency questionnaire (FFQ),  referring to habitual consumption in the two years preceding cancer diagnosis (cases) or hospital admission (controls)~\citep{franceschi_reproducibility_1993}. 
Specifically, three distinct studies contributed the data: 
(i) an oral cavity and pharyngeal cancer study (1991–2009; Pordenone, Milan, Latina; $n=2437$)~\citep{edefonti2022diet}; (ii) a laryngeal cancer study (1992–2000; Pordenone, Gorizia, Udine, Milan; $n=1322$)~\citep{accardi_dietary_2019}; and (iii) an esophageal cancer study (1992–1997; Milan, Pordenone, Gorizia, Padua; $n=1047$)~\citep{bravi_dietary_2012}.

We post-processed the 78 FFQ responses measured as raw frequencies per week of foods or composite recipes in prespecified standard or natural portion sizes (see Supplementary Materials for details). We focused
on the food items with the highest correlation (Spearman correlation $\geq 0.2$ with at least one other food item), resulting in a final set of $K=38$ food items.
Covariates included in the probit part of the model were sex, age, education level, and total energy intake (kcal/day). Duplicate control subjects across studies, individuals with missing values on food items and those with implausible total energy intake (i.e., $<$ 1-st percentile or $>$ 99-th percentile) were removed. This resulted in a final sample of 1652 cancer cases and 3390 controls. Table~S1 in the Supplementary Materials summarizes the socio-demographic characteristics by case-control status in the three studies.

We ran the ziNMF with hyperparameters set to $L=5$, $\alpha^w=1$, $\beta^w=25$, $c_{kl}=10$, $\theta_{k1}=0.5$, $\gamma_1^\theta=1.5$, $\gamma_2^\theta=20$, $\alpha^m=\beta^m=1$, $\mathbf{\beta}_0=(1.5, 0,0,0,0)$, and $\tau_0=5$.  We initialize the Gibbs sampler with $R=15$ initial patterns, converging to 14 estimated dietary patterns.  

\begin{figure}[ht!]
    \centering
    \includegraphics[width=\linewidth]{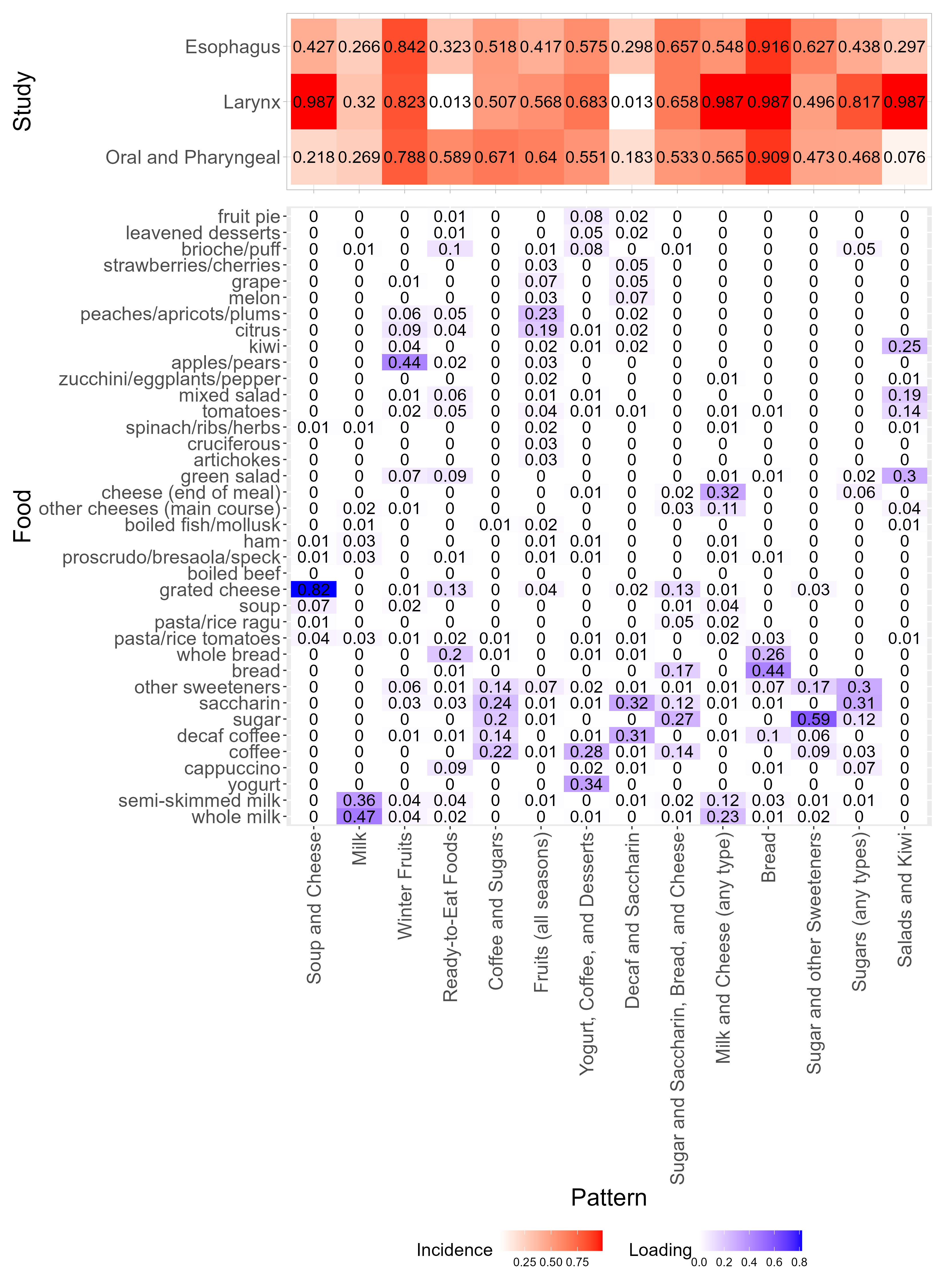}
    \caption{Estimated study-level incidence and posterior median loadings for each of the $R=14$ estimated dietary patterns $\mathbf{W}$. For visualization purposes, loadings were scaled to sum to 1  within each pattern. Pattern Incidence was inferred using ziNMF, defined as the proportion of subjects assigned to cluster 2 or higher under the dependent Dirichlet process.}\label{fig:loadings}
\end{figure}

We first present the estimated dietary patterns, i.e., the posterior median of the loading matrix $\mathbf{W}$ (Figure~\ref{fig:loadings}). Dietary patterns were labeled based on dominant (i.e., $> 0.07$) loadings and include: (1) Soup and Cheese, (2) Milk, (3) Winter Fruits, (4) Ready-to-Eat Foods, (5) Coffee and Sugars, (6) Fruits (all seasons), (7) Yogurt, Coffee, and Desserts, (8) Decaf and Saccharin, and (9)Sugar and Saccharin, Bread and Cheese. (10) Milk and Cheese (any type), (11) Bread, (12) Sugar and other Sweeteners, (13) Sugars (any types), and (14) Salads and Kiwi. 

The top panel of Figure~\ref{fig:loadings} shows the estimated study-level prevalence of each dietary pattern across the three cancer studies: Oral and Pharyngeal, Laryngeal, and Esophageal.  We define study-level prevalence as the proportion of individuals assigned to cluster 2 or higher under the DDP, indicating moderate to high pattern exposure to the dietary pattern (cluster 1 corresponds to low consumers).

Several dietary patterns were consistently prevalent across all studies. For example, the Bread pattern is highly prevalent in all three cancer types, with incidence rates of 91.6\%, 98.7\%, and 90.9\%, in the oral and pharyngeal, laryngeal and esophageal cancer studies, respectively. Other patterns that were present in the majority of subjects in each of the studies included: Winter Fruits; Coffee and Sugars; Yogurt, Coffee, and Desserts; Sugar and Saccharin, Bread, and Cheese; and Milk and Cheese (any type).

In contrast, some patterns showed heterogeneity across studies. The Ready-to-Eat Foods appeared in 58.9\% of participants in the oral and pharyngeal cancer study but only 1.3\% in the laryngeal cancer study. Similarly, the Decaf and Saccharin patterns was present in 29.8\%  of the esophageal cancer cohort but only 1.3\% of the laryngeal cohort. 
 The Salads and Kiwi pattern demonstrated comparable disparity, appearing in just 7.6\% of participants in the oral and pharyngeal study but in 98.7\% of those in the laryngeal study.

These results suggest important study-specific dietary variation and highlight the need to quantify dietary pattern prevalence as a continuous measure. Rather than assuming patterns are either universally shared or entirely study-specific, ziNMF enables a nuanced characterization of dietary behavior heterogeneity, critical for identifying differential dietary risk factors across populations.

\begin{figure}
    \centering
    \includegraphics[width=0.5\linewidth]{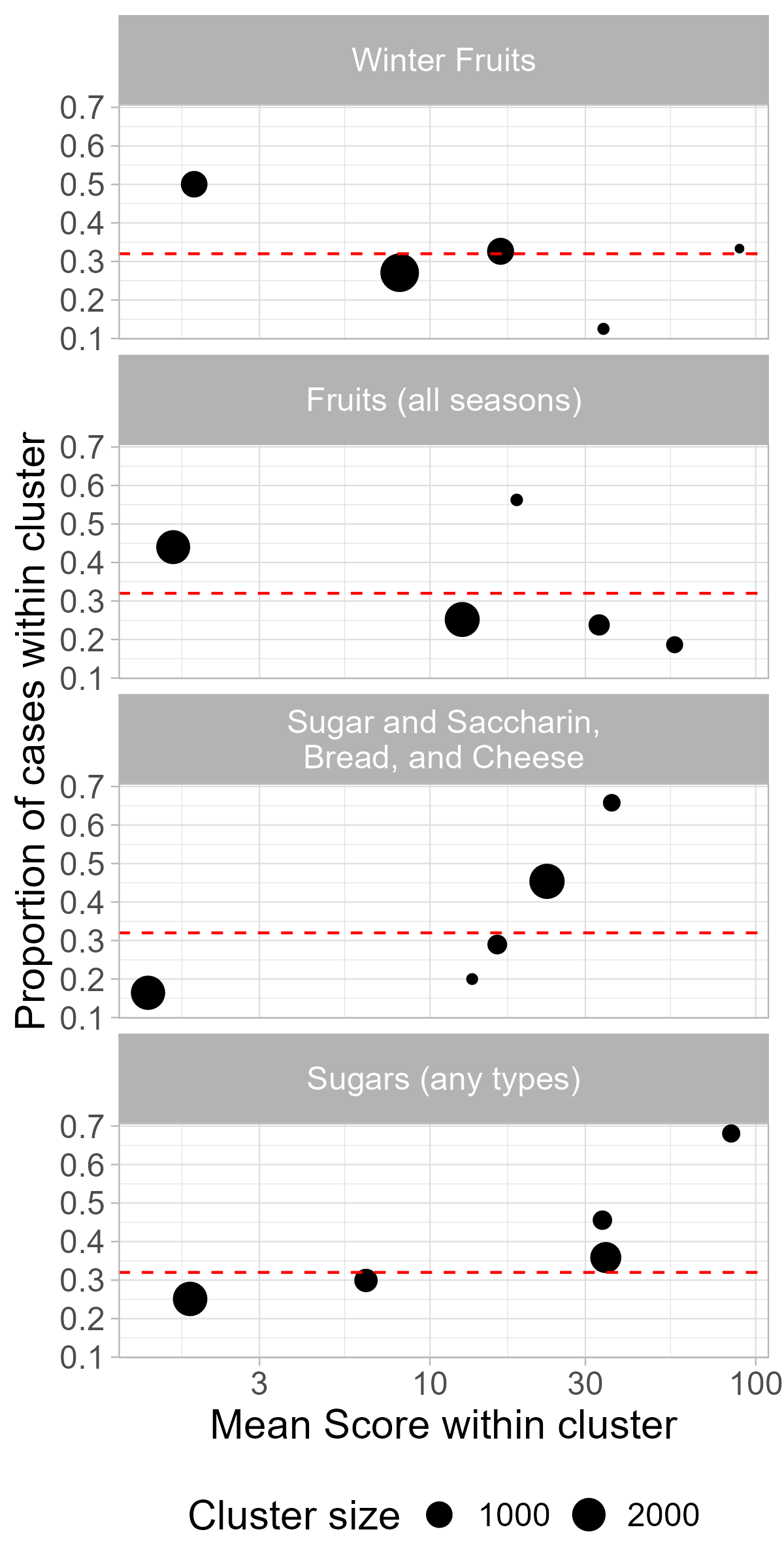}
    \caption{Exploratory associations between dietary patterns and cancer, based on cluster assignments from the dependent Dirichlet process. Each point represents a cluster, with the x-axis indicating the mean pattern score, the y-axis the proportion of cancer cases, and the point size corresponding to the number of individuals in the cluster. The horizontal dashed lines indicate the overall sample prevalence of cancer (32\%).}
    \label{fig:exploratory}
\end{figure}

To explore the associations between dietary patterns and cancer risk, we cluster individuals based on their estimated pattern scores (see Supplementary Materials, Section 2.1) and visualize the resulting clusters based on the mean dietary pattern scores, proportion of cancer cases, and cluster size. Figure~\ref{fig:exploratory} displays selected dietary patterns that showed notable associations with cancer: Winter Fruits; Fruits (all seasons); Sugar and Saccharin, Bread and Cheese; and Sugars (any type). All 14 patterns are shown in Supplementary Figure~S3. We find that the Winter Fruits and Fruits (all seasons) were negatively associated with cancer: clusters with low pattern scores had 15-20\% more cases than the overall sample prevalence (i.e., 32\%), while clusters with higher pattern scores contained 10-20\% fewer cases than the same baseline. Conversely, the Sugar and Saccharin, Bread and Cheese and Sugars (any types) patterns exhibited positive associations: clusters with low pattern scores had 5-15\% fewer cases than baseline, while clusters with very high pattern scores had as much as 40\% more cases than baseline. These findings are consistent with previous literatureindicating that high consumption of sugar-sweetened beverage are associated with oral cancer~\citep{gomez2025high}, while high-fiber diet have been linked to reduced cancer incidence.

These exploratory findings suggest that ziNMF not only  captures dietary patterns across studies but alsoreveals meaningful signal that could drive insights into disease associations.

\FloatBarrier

\section{Discussion}\label{sec:Discussion}

In this paper, we introduce a novel Bayesian zero-inflated multi-study NMF (ziNMF) model that offers several key methodological advancements. First, the model is explicitly designed for multi-study analysis, enabling the identification of both shared and study-specific structure. Second, by incorporating a point mass at zero within the mixture distribution, our model directly accounts for excess zeros, substantially improving estimation accuracy relative to existing multi-study NMF methods, as demonstrated in simulation studies benchmarking against Compressive NMF~\citep{zito2024compressive} and Bayesian Multi-Study NMF~\citep{grabski2025bayesian}. Third, leveraging a BNP framework, our model induces flexible clustering of individuals based on their latent pattern scores, providing a flexible and principled tool for complex and heterogeneous datasets. 

In nutritional epidemiology, individuals are typically grouped based on their consumption levels using quantile-based categorization (e.g., tertiles). This approach has several advantages, most notably that the lowest consumption group can serve as a natural reference category when estimating disease risk in regression models~\citep{schulze2018food}. However, such categorization does not account for potential confounding factors. In contrast, our framework allows for clustering individuals based not only on their dietary pattern scores but also on covariate information incorporated through the prior. This enables us to account for confounding and recover a more accurate and interpretable signal. Evidence supporting this advantage is presented in the Supplementary Materials, where we compare the ability of our method to recover the true signal against a traditional two-step approach using NMF followed by tertile categorization (Figure S1).

We demonstrate that in simulated settings, the clustering provided by our method is more accurate than the typical approach to cluster subjects based on quantiles. This makes our method particularly well suited to areas such as nutritional epidemiology, where data are often sparse and integrated across multiple studies. However, the model is also generalizable to other settings, such as genomics \citep{grabski2025bayesian} and ecology \citep{stolf2024infinite}, where the goal is to jointly analyze multiple count-based studies with excess zeros and to cluster subjects into meaningful groups.

We illustrate the effectiveness of our method through an application to a nutritional epidemiology dataset characterized by excess zeros and multi-study structure induced by different cancer cases.
Overall, we find that the dietary patterns and clusters learned by ziNMF not only provide accurate decompositions of the observed data, but capture relevant epidemiological signal which may be of interest to investigators. Additionally, our method has the added benefit of clustering subjects while simultaneously learning dietary patterns, which is more statistically principled and flexible compared to previous work in nutritional epidemiology \citep{edefonti2009clustering}, allowing for intuitive model based association between the dietary patterns and cancer.

Several important challenges in multi-study NMF remain beyond the scope of this work and offer promising directions for future research. One such challenge is the well-known issue of identifiability of the NMF model.
While some frequentist approaches address unidentifiability via constrained optimization under specific assumptions~\citep{fu_robust_2016}, there is a need for fully Bayesian formulations that guarantee identifiability through appropriate prior specifications or model reparameterizations. A second limitation concerns our choice of independent gamma priors for the elements of the loading matrix $\mathbf{W}$, which introduces scale non-identifiability between loadings and scores. While alternative priors, such as the Dirichlet distribution~\citep{zito2024compressive, hansen2025bayesian}, can address this issue by normalizing the columns of $\mathbf{W}$ to sum to 1, we opted for gamma priors to preserve conjugacy with the zero-inflated Poisson likelihood and ensure computational efficiency. 
Further work is needed to explore trade-offs between identifiability, flexibility, and computational tractability in this context.

We hope that our proposed framework will encourage integrated analyses across multiple studies and help address longstanding challenges related to the replicability, generalizability, and robustness of unsupervised learning methods, both within nutritional epidemiology and across the broader data science.


\backmatter


\section*{Acknowledgements}

The authors thank Carlo La Vecchia, Werner Garavello, and Luigino Dal Maso for providing the data used in the cancer case study application and Arianna Natale for the preprocessing of food items and covariates. RDV was supported by NIH grant P20GM109035 (COBRE CBHD) and the Italian Ministry of Education (Rita Levi Montalcini program). VE was supported by the Italian Ministry of University and Research grant PRIN 20227YCB5P.
\vspace*{-8pt}






\bibliographystyle{biom} \bibliography{biomsample}

\appendix





\label{lastpage}

\end{document}


\maketitle

\section{Posterior Updates}

Recall the model:
\begin{align}
    m_{pis} &\sim  \pi_{ps} \delta_0  + \left(1-\pi_{ps}\right)\text{Poisson}\left(\sum_{k=1}^K w_{pk}h_{kis}\right) \label{obs_like},
\end{align}
where $m_{pis} \in \mathcal{Z}_+$ is a poisson random variable with mean $\sum_{k=1}^K w_{pk}h_{kis}$ and additional mass at $0$ given by $\pi_{ps}$. For the purposes of sampling, we augment model~\eqref{obs_like} with binary indicator $a_{pis}$ with full hierarchical representation:
\begin{align}
     m_{pis} \mid a_{pis} &\sim  a_{pis} \delta_0  + \left(1-a_{pis}\right)\text{Poisson}\left(\sum_{k=1}^K w_{pk}h_{kis}\right), \\
     a_{pis} \mid \pi_{ps} &\sim \text{Bernoulli}(\pi_{ps}), \\
     \pi_{ps} &\sim \text{Beta}(\alpha^m, \beta^m),
\end{align}
where $\alpha^m, \beta^m \in \mathcal{R}^2_+$ are considered fixed hyper-parameters. Additionally, we augment the model with latent counts $z_{pkis}$ such that $\sum_{k}z_{pkis}=m_{pis}$, 
\begin{align}
   z_{pkis} \mid a_{pis} &\sim a_{pis}\delta_0 + (1-a_{pis}) \text{Poisson}\left(w_{pk}h_{kis}\right),
\end{align}
which combined give the complete data-likelihood:
\begin{align}
    \prod_{s=1}^S \prod_{i=1}^{N_s} \prod_{p=1}^P \prod_{k=1}^K  \mathbbm{1}\left\{ z_{pkis}=0 \right\}^{a_{pis}}\left((w_{pk}h_{kis})^{z_{pkis}}\exp\left\{-w_{pk}h_{kis}\right\}/z_{pkis}!\right)^{1-a_{pis}},
\end{align}
with the constraint $m_{spi} = \sum_{k} z_{pkis}$.

\subsection{Latent Counts}
By construction, the latent counts $z_{p1is}= \dots=z_{pKis}=0$ when $a_{spi}=1$. When $a_{spi}=0$, they are independently Poisson distributed with known sum $\sum_{k} z_{pkis} = m_{pis}$. Therefore, we have:
\begin{align*}
    z_{p1is}, \dots, z_{pKis} \mid a_{spi}=0,  \dots &\sim \text{Multinomial}\left(m_{pis}, \boldsymbol{\pi}^z_{pis}\right), \\
    \pi^z_{pkis} &\propto w_{pk}h_{kis},
\end{align*}
which allows us to sample $z_{p1is}, \dots, z_{pKis}$ conditional on indicators $\mathbf{A}_s$, pattern matrix $\mathbf{W}$ and scores matrix $\mathbf{H}_s$.

\subsection{Excess Zero Indicators}
The conditional probabilities that $a_{api}=1$, $a_{api}=0$, are given by:
\begin{align*}
    P(a_{spi}=1\mid\dots) &\propto p(a_{spi}=1)p(m_{pis}\mid a_{spi}=1, \dots), \\
    &= \pi_{ps}\mathbbm{1}\left\{m_{pis}=0\right\} , \\
    P(a_{spi}=0\mid\dots) &\propto p(a_{spi}=0)p(m_{pis}\mid a_{spi}=0,\dots), \\
    &= (1-\pi_{ps}) \text{Poisson}\left(m_{pis}, \sum_{k=1}^K w_{pk}h_{kis}\right),
\end{align*}
where $\text{Poisson}(\cdot, \lambda)$ denotes the Poisson density with parameter $\lambda$. Therefore,
\begin{align*}
    a_{spi} \mid \dots &\sim \text{Bernoulli}\left(\frac{P(a_{spi}=1\mid\dots)}{P(a_{spi}=1\mid\dots) + P(a_{spi}=0\mid\dots)}\right).
\end{align*}

\subsection{Excess Zero Probabilities}
The conditional posterior for $\pi_{ps}$ is given by:
\begin{align*}
    p(\pi_{ps} \mid \dots) &\propto p(\pi_{ps})p(\dots \mid \pi_{ps}), \\
    &\propto (\pi_{ps})^{\alpha^m - 1}(1-\pi_{ps})^{\beta^m - 1} \prod_{i=1}^{N_s} (\pi_{ps})^{a_{pis}}(1-\pi_{ps})^{1-a_{pis}}, \\
    &= \propto (\pi_{ps})^{\alpha^m + \sum_{i=1}^{N_s}a_{pis}- 1}(1-\pi_{ps})^{\beta^m +N_s - \sum_{i=1}^{N_s}a_{pis} - 1},
\end{align*}
which implies:
\begin{align*}
    \pi_{ps} \mid \dots &\sim \text{Beta}\left(\alpha^m + \sum_{i=1}^{N_s}a_{pis}, \beta^m +N_s - \sum_{i=1}^{N_s}a_{pis}\right).
\end{align*}

\subsection{Pattern Matrix}
The posterior distribution for the loading $p$ of pattern $k$ is given by:
\begin{align*}
    p(w_{pk}\mid \dots) &\propto p(w_{pk})p(\dots\mid w_{pk} ), \\
    &\propto  (w_{pk})^{\alpha^w-1} \exp\left\{-w_{pk}\beta\right\} \prod_{s=1}^S\prod_{i=1}^{N_s} (w_{pk})^{z_{pkis}}\exp\left\{-w_{pk} h_{ski}(1-a_{pis})\right\}, \\
    &= \left(w_{pk}\right)^{\alpha^w  +\sum_{s}\sum_i z_{pkis} -1}\exp\left\{-w_{pk} \left(\beta^w +\sum_{s}\sum_i h_{ski}(1-a_{pis}) \right)\right\},
\end{align*}
which implies:
\begin{align*}
    w_{pk} &\sim \text{Gamma}\left(\alpha^w  +\sum_{s}\sum_i z_{pkis}, \beta^w +  \sum_{s}\sum_i h_{ski}(1-a_{pis}) \right).
\end{align*}

\subsection{Scores Matrix}
Following \cite{rodriguez2011nonparametric}, it has been shown that that a finite truncation of the dirichlet probit stick-breaking process is a good approximation. Setting $L^*$ as an appropriate upper bound, the prior for the score of pattern $k$ in individual $i$ of study $s$ is given by:
\begin{align*}
     p(h_{kis}\mid \boldsymbol{\omega}_{kis}, \mathbf{c} , \boldsymbol{\theta}_{k}) &= \sum_{l=1}^{L^*} \omega_{klis}\left(\mathbf{x}_{is}\right)\text{Gamma}\left( c_l, c_l/\theta_{kl} \right), \\
    \omega_{klis}\left(\mathbf{x}_{is}\right)&= \Phi\left(\alpha_{klis}(\mathbf{x}_{is})\right)\prod_{r<l}\left\{\Phi\left(1-\alpha_{kris}(\mathbf{x}_{is})\right)\right\}, \\
    \alpha_{klis}(\mathbf{x}_{is})\mid  \boldsymbol{\beta}_{kls} &\sim \mathcal{N}\left(\mathbf{x}_{is}^\intercal \boldsymbol{\beta}_{kls},1\right), \\
    \boldsymbol{\beta}_{kls} &\sim \mathcal{N}\left(\boldsymbol{\beta}_{l}^{0}, (\tau^{0})^{-1}\text{I}_{Q_s}\right), \\
    c_l &= \begin{cases}
        1 & l=1, \\
        c & l\geq 2,
    \end{cases} \\
    \theta_{kl} &\sim \begin{cases}
        \delta_\epsilon & l=1, \\
        \text{Inv-Gamma}\left(\gamma_1^\theta, \gamma^\theta_2\right) & l \geq 2,
    \end{cases}
\end{align*}
where $\mathbf{c} = (c_1,\dots,c_{L^*})^\intercal$, $\boldsymbol{\theta}_{k}=(\theta_{k1},\dots, \theta_{kL^*})^\intercal$, and $\omega_{kL^*is}\left(\mathbf{x}_{is}\right)=1$. To sample from the posterior distribution of $h_{kis}$, we augment with latent binary vector $\mathbf{d}_{kis}$ of dimension $L^*$, where $d_{klis}=1$ if $h_{kis}$ belongs to cluster $l$, and $0$ otherwise. We can rewrite the prior as the following:
\begin{align*}
    p(h_{kis}\mid \mathbf{d}_{kis}, \mathbf{c} , \boldsymbol{\theta}_{k}) &= \prod_{l=1}^{L^*} \text{Gamma}\left(c_l, c_l/\theta_{kl}\right)^{d_{klis}}, \\
    \mathbf{d}_{kis} \mid  \boldsymbol{\omega}_{kis}\left(\mathbf{x}_{is}\right) &\sim \text{Multinomial}(1, (\omega_{k1is}\left(\mathbf{x}_{is}\right),\dots,\omega_{kL^*is}\left(\mathbf{x}_{is}\right))).
\end{align*}

Therefore, the posterior distribution of $h_{ski}$ is given by:
\begin{align*}
    p(h_{kis} \mid  \dots) &\propto p(h_{kis}\mid  \mathbf{d}_{kis}, \mathbf{c} , \boldsymbol{\theta}_{k}) )p(\dots \mid  h_{kis}) \\
    &=  \prod_{l=1}^{L^*} \left( (h_{kis})^{c_l - 1} \exp\left\{-h_{kis} c_l/\theta_{kl} \right\} \right)^{d_{klis}} \prod_{p=1}^P \left((h_{kis})^{z_{pkis}}\exp\left\{-w_{pk}h_{kis}\right\}\right)^{1-a_{pis}} \\
    &= (h_{kis})^{c_{kis} + \sum_{p=1}^P z_{pkis}(1-a_{pis})}\exp\left\{-h_{kis} \left(c_{kis}/\theta_{kis} + \sum_{p=1}^P w_{pk}(1-a_{pis})\right) \right\},
\end{align*}
which implies:
\begin{align*}
    h_{kis} \mid  \dots &\sim \text{Gamma}\left(c_{kis} + \sum_{p=1}^P z_{pkis}(1-a_{pis}), c_{kis}/\theta_{kis} + \sum_{p=1}^P w_{pk}(1-a_{pis})\right),
\end{align*}
where $c_{kis}= \mathbf{c}^\intercal\mathbf{d}_{kis}$  and $\theta_{kis}=\boldsymbol{\theta}_{k}^\intercal\mathbf{d}_{kis}$ are equal to $c_l$ and $\theta_{kl}$ for the cluster corresponding to $h_{kis}$.

\subsubsection{Cluster Indicators}
The posterior distribution of $\mathbf{d}_{kis}$ is given by:
\begin{align*}
    p(\mathbf{d}_{kis}\mid \dots) &\propto p(\mathbf{d}_{kis})p(\dots\mid\mathbf{d}_{kis}), \\
    &= \prod_{l=1}^{L^*} (\omega_{klis})^{d_{klis}}\left(\text{Gamma}\left(h_{kis}\mid c_l, \theta_{kl}\right)\right)^{d_{klis}},
\end{align*}
which implies the posterior of $\mathbf{d}_{kis}$ is a multinomial distribution with probabilities:
\begin{align*}
    P(d_{klis}=1) &\propto \omega_{klis}\text{Gamma}\left(h_{kis}\mid c_l, \theta_{}\right).
\end{align*}

\subsubsection{Cluster Means}
The posterior distribution for cluster means $\theta_{kl}$ for $l=2,\dots,L^*$ is given by:
\begin{align*}
    p(\theta_{kl}\mid \dots) &\propto p(\theta_{kl})p(\dots\mid\theta_{kl}), \\
    &\propto (\theta_{kl})^{-\gamma_1^\theta-1}\exp\left\{-\gamma^\theta_2/\theta_{kl}\right\}\prod_{s=1}^S\prod_{i=1}^{N_s} \left((c_l/\theta_{kl})^{c_l}\exp\left\{-c_l/\theta_{kl}h_{kis}\right\}\right)^{d_{klis}}, \\
    &= \left(\theta_{kl}\right)^{-(\gamma_1^\theta +c_l\sum_{s=1}^S\sum_{i=1}^{N_s}d_{klis}) - 1}\exp\left\{-\left(\gamma_2^\theta + c_l\sum_{s=1}^S\sum_{i=1}^{N_s} h_{kis}d_{klis}\right)/\theta_{kl}\right\},
\end{align*}
which implies:
\begin{align*}
    \theta_{kl}\mid \dots &\sim \text{Inv-Gamma}\left(\gamma_1^\theta + c_l\sum_{s=1}^S\sum_{i=1}^{N_s}d_{klis}, \gamma_2^\theta + c_l\sum_{s=1}^S\sum_{i=1}^{N_s} h_{kis}d_{klis}\right).
\end{align*}

\subsubsection{Probits}
To sample weights $\omega_{klis}$ and covariate effects $\beta_{kls}$, we need to perform one last data augmentation. Following \cite{rodriguez2011nonparametric}, we sample the augmented variables $y_{klis}$ from its full conditional distribution for $l=1,\dots,L^*-1$:
\begin{align*}
    y_{klis} \begin{cases}
        \mathcal{N}_+\left(\mathbf{x}_{is}^\intercal \boldsymbol{\beta}_{kls},1\right) & d_{klis} = 1, \\
         \mathcal{N}_-\left(\mathbf{x}_{is}^\intercal \boldsymbol{\beta}_{kls},1\right) & \sum_{r=1}^{l} d_{kris} = 0,
    \end{cases}
\end{align*}
where $\mathcal{N}_+(\mu, 1)$, $\mathcal{N}_{-}(\mu, 1)$ denotes the normal distribution truncated above or below zero, respectively, allowing us to sample positive $y_{klis}$ for $h_{kis}$ belonging to cluster $l$ and sample negative $y_{klis}$ for $h_{kis}$ belonging to clusters $l+1,\dots, L^*$, and $y_{klis}$ for $h_{kis}$ belonging to clusters $1,\dots,l-1$ is undefined.

\subsubsection{Covariate Effects}
Lastly, the posterior distribution for covariate effects $\beta_{kls}$ for $l=1,\dots,L^*-1$ is given by:
\begin{align*}
    \beta_{kls} &\sim \mathcal{N}\left(\boldsymbol{\mu}_{kls}, \boldsymbol{\Sigma}_{kls}\right), \\
    \boldsymbol{\Sigma}_{kls} &= \left(\mathcal{\tau^0I_{Q_s}} + \tilde{\mathbf{x}}_{s}^\intercal\tilde{\mathbf{x}}_{s}\right)^{-1}, \\
    \boldsymbol{\mu}_{kls} &= \boldsymbol{\Sigma}_{kls}\left(\tau^0I_{Q_s}\boldsymbol{\beta}^0_l + \tilde{\mathbf{x}}_{s}^\intercal\tilde{\mathbf{y}}_{s}\right),
\end{align*}
where $\tilde{\mathbf{x}}_{s}$, $\tilde{\mathbf{y}}_{s}$ are constructed from $\mathbf{x}_{s}$, $\mathbf{y}_{s}$ to contain individuals such that the corresponding score $h_{ski}$ belongs to cluster $l$ or higher, i.e.:  $\tilde{\mathbf{x}}_{s}=\left\{\mathbf{x}_{is} : \sum_{r=l}^{L^*} d_{kris}=1  \right\}$, $\tilde{\mathbf{y}}_{s}=\left\{\mathbf{y}_{is} : \sum_{r=l}^{L^*} d_{kris}=1  \right\}$.

\section{Additional Simulation Results}
\subsection{Clustering}\label{sec:sim_rand_indx}
To assess the ability of ziNMF to recover latent clusters, we modify Scenarios 1 and 2 such that Pattern 1 is expressed via three underlying clusters, with Gamma-distributed scores: $(1, 0.25)$, $(10, 1)$, and $(10, 0.2)$. Cluster membership depends on subject-specific covariates. 

To estimate clusters, we first fit ziNMF on the simulated data. Then, we calculate point estimates for cluster membership based on posterior samples of $\mathbf{d}_{1is}$ using the \texttt{BNPmix} R package~\citep{BNPmixPackage}, which finds the partition that minimizes a loss function based on the variation of information \citep{wade_bayesian_2018}. We compare the labels provided by ziNMF with a two-step approach based on previous work in nutritional epidemiology: first dietary patterns are estimated via NMF (i.e.,  CompNMF) and then subjects are grouped based on the tertile of the estimated pattern scores, which we label ``CompNMF + Tertile''. 

\begin{figure}[!ht]
    \centering
    \includegraphics[width=0.5\linewidth]{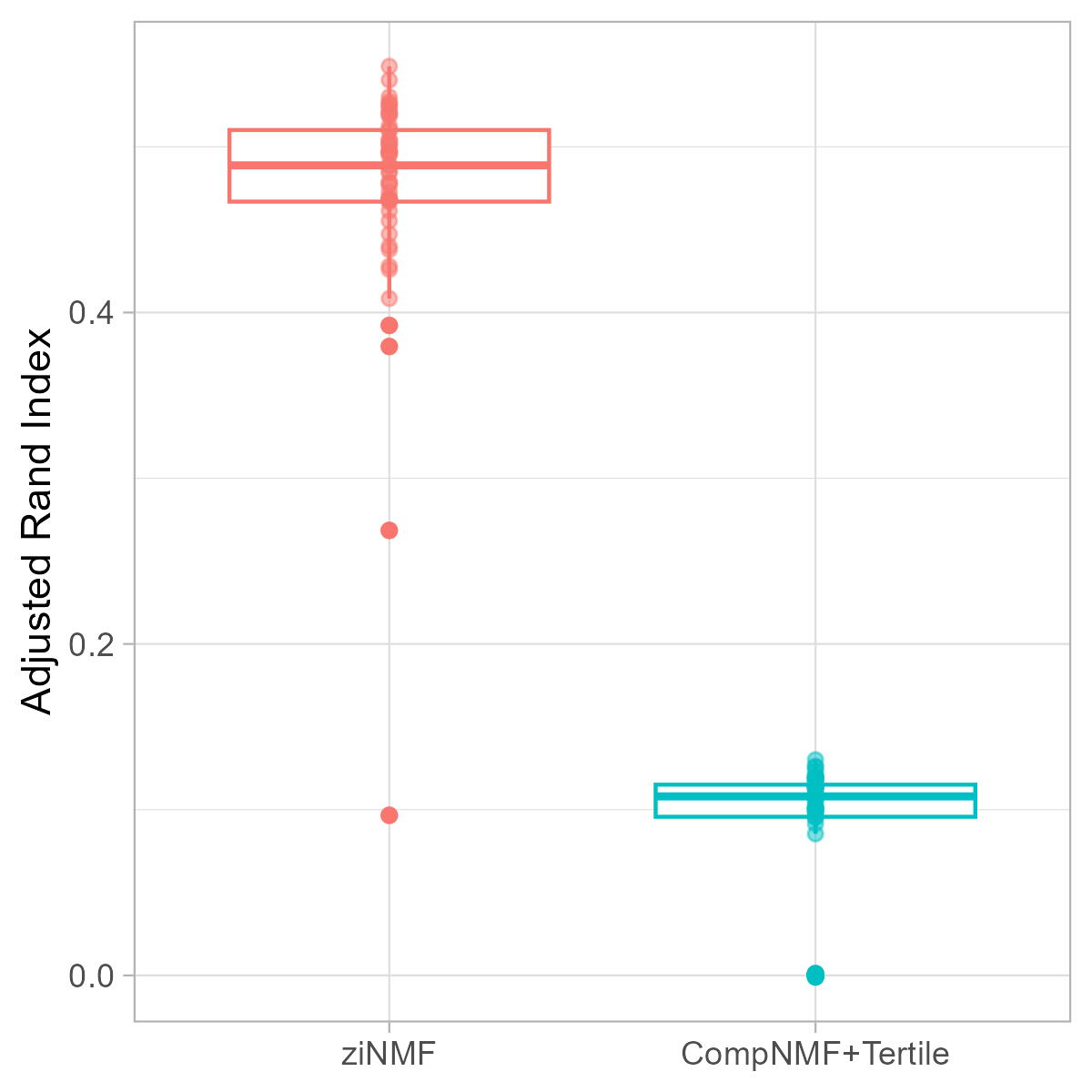}
    \caption{Adjusted Rand Index for ziNMF and CompNMF on a simulation based on modification of Scenario 1 and 2.}\label{fig:rand-idx}
\end{figure}

We evaluate clustering accuracy using the adjusted Rand index (Figure~\ref{fig:rand-idx}). The adjusted rand index is a metric which assesses the alignment of two proposed partitions ranging from -1 to 1, where values near 1 indicate perfect alignment and values near 0 indicate that any similarity between the two partitions is due to random chance.
Our ziNMF model achieves an average adjusted Rand index just below 0.5, significantly outperforming CompNMF + Tertile approach, which averages near 0.1. These results further support that ziNMF provides substantially better cluster recovery in settings with heterogeneous latent structure.

\section{Cancer Case Study Data Analysis}

We postprocess the dietary data for analysis,  responses from the Food Frequency Questionnaire (FFQ) (i.e., raw frequencies per week of foods or composite recipes in prespecified standard or natural portion sizes). The majority of the 78 food items (ali) were originally recorded as counts, only 12 items---primarily fruits, vegetables, and ice cream----were adjusted for seasonality by multiplying their reported frequency by the fraction of the year in which they were consumed (e.g., 2/12 for items eaten during two months). Occasional consumption (0.5) was recoded as zero. Reported portion sizes were standardized by multiplying frequencies by 0.7 for small portions and 1.3 for large portions; however, the vast majority of individuals reported medium portions, meaning that no adjustment or rounding was necessary in most cases.

As a result, only a small subset of individuals required rounding, primarily those consuming seasonal items or reporting non-average portion sizes. All fractional values were then rounded to the nearest whole number to yield count data suitable for statistical modeling.

Table~\ref{tab:the-studies} summarizes demographic and lifestyle characteristics by case-control status across the three cancer types. Age and total energy intake were generally comparable between cases and controls. However, sex distribution differed, particularly in the Oral and Pharyngeal study, where 70\% of cases were male compared to 60\% of controls. Smaller discrepancies were also observed in the Larynx (82\% vs. 76\%) and Esophagus (73\% vs. 66\%) studies. Education was largely balanced, except in the Larynx study, where a smaller proportion of cases had less than 8 years of education (52\% vs. 62\%).

\begin{table}[!ht]
\centering\begingroup\fontsize{6}{9}\selectfont
\caption{Summary of the case study data. For each study, we report the sample size, the number of cancer cases, and distributions of key demographic and lifestyle variables, including sex, age, years of education, and total energy intake (kcal/day).  For categorical variables, we present absolute frequencies and corresponding percentage; for continuous  variables, we report medians and the interquartile range.}\label{tab:the-studies} 
\centering
\begin{tabular}{lcccccc}
 & \multicolumn{2}{c}{Oral and Pharynx} & \multicolumn{2}{c}{Larynx} & \multicolumn{2}{c}{Esophagus} \\
Study & Cases & Controls & Cases & Controls & Cases & Controls \\
\toprule N & 909 & 2402 & 446 & 622 & 297 & 366\\
Sex & 727(80\%) & 1444(60\%) & 403(90\%) & 488(78\%) & 268(90\%) & 284(78\%)\\
Age & 58(52-66) & 58(50-66) & 61(55-67) & 61(55-68) & 60(55-66) & 60(53-67)\\
Education years $<$ 9 & 746(82\%) & 1783(74\%) & 368(83\%) & 500(80\%) & 262(88\%) & 280(77\%)\\
Education years $>$ 13 & 37(4\%) & 138(6\%) & 22(5\%) & 28(5\%) & 5(2\%) & 25(7\%)\\
Total energy intake & 2557(2069-3249) & 2330(1902-2832) & 3018(2476-3549) & 2477(2019-3043) & 2792(2185-3330) & 2335(1855-2974)\\
\bottomrule
\end{tabular}
\endgroup{}
\end{table}

The observed proportion of zeros for each food item in each study is reported in Table~\ref{tab:nonconsume_rates}.
We observed high variability in zero-inflation across food items and studies, with overall non-consumption rates ranging from 7.6\% to 98\% (Table~\ref{tab:nonconsume_rates}).  The most frequently unconsumed items within study included decaf coffee (98\%), saccharin (97\%), other sweeteners (95\%) cappuccino (90\%) and leavened desserts (90\%). Food items with the lowest percentage of zeros were: bread (7.6\%), green salad (9.9\%), coffee (11\%), and apples/pears (16\%).

\begin{table}[!ht]
\centering\begingroup\fontsize{6}{8}\selectfont
\caption{Observed non-consumption rates of the $P=28$ selected food items across the $S=3$ studies. Rates are calculated as the proportion of subjects reporting zero consumption for each item.}\label{tab:nonconsume_rates}
\centering
\begin{tabular}[t]{lrrr}
\toprule
  & Oral and Pharyngeal & Larynx & Esophagus\\
\midrule
whole milk & 0.700 & 0.710 & 0.740\\
semi-skimmed milk & 0.730 & 0.720 & 0.760\\
yogurt & 0.750 & 0.800 & 0.830\\
cappuccino & 0.870 & 0.900 & 0.900\\
coffee & 0.120 & 0.120 & 0.110\\
decaf coffee & 0.950 & 0.950 & 0.980\\
sugar & 0.180 & 0.180 & 0.170\\
saccharin & 0.970 & 0.970 & 0.970\\
other sweeteners & 0.950 & 0.950 & 0.950\\
bread & 0.076 & 0.073 & 0.081\\
whole bread & 0.930 & 0.930 & 0.930\\
pasta/rice tomatoes & 0.200 & 0.260 & 0.310\\
pasta/rice ragu & 0.470 & 0.340 & 0.500\\
soup & 0.400 & 0.410 & 0.380\\
grated cheese & 0.220 & 0.160 & 0.200\\
boiled beef & 0.780 & 0.710 & 0.680\\
proscrudo/bresaola/speck & 0.380 & 0.440 & 0.500\\
ham & 0.570 & 0.610 & 0.630\\
boiled fish/mollusk & 0.470 & 0.560 & 0.630\\
other cheeses (main course) & 0.400 & 0.370 & 0.410\\
cheese (end of meal) & 0.600 & 0.420 & 0.640\\
green salad & 0.150 & 0.099 & 0.140\\
artichokes & 0.750 & 0.870 & 0.800\\
cruciferous & 0.680 & 0.810 & 0.800\\
spinach/ribs/herbs & 0.570 & 0.690 & 0.710\\
tomatoes & 0.180 & 0.200 & 0.250\\
mixed salad & 0.570 & 0.480 & 0.790\\
zucchini/eggplants/pepper & 0.500 & 0.520 & 0.640\\
apples/pears & 0.160 & 0.170 & 0.220\\
kiwi & 0.700 & 0.790 & 0.810\\
citrus & 0.230 & 0.280 & 0.300\\
peaches/apricots/plums & 0.180 & 0.230 & 0.290\\
melon & 0.720 & 0.820 & 0.810\\
grape & 0.570 & 0.610 & 0.650\\
strawberries/cherries & 0.790 & 0.890 & 0.880\\
brioche/puff & 0.800 & 0.870 & 0.890\\
leavened desserts & 0.870 & 0.900 & 0.920\\
fruit pie & 0.820 & 0.780 & 0.870\\
\bottomrule
\end{tabular}
\endgroup{}
\end{table}

\begin{figure}[ht!]
    \centering
    \includegraphics[width=\linewidth]{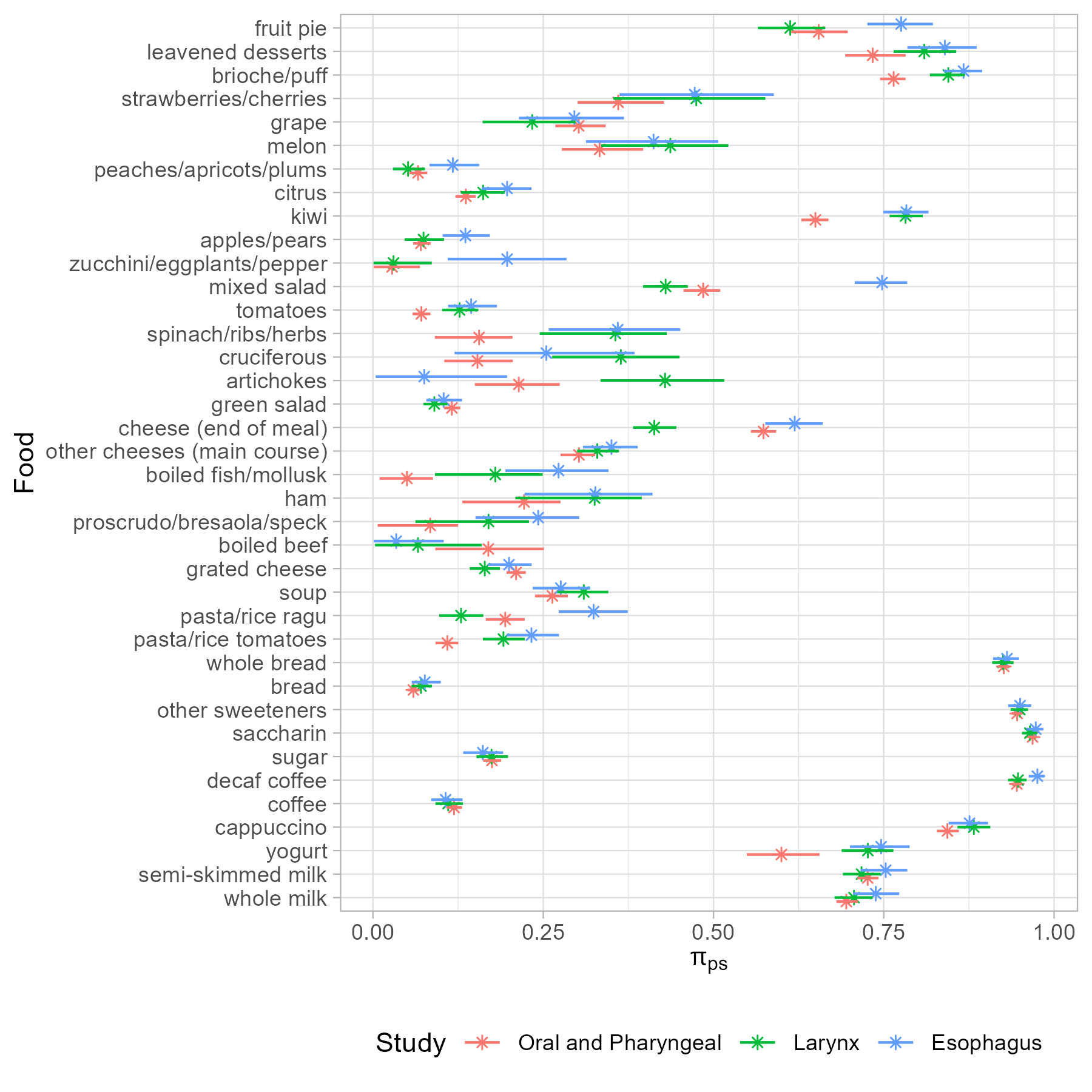}
    \caption{Estimated study-specific excess zero probabilities $\pi_{ps}$ and corresponding 95\% credible intervals.}\label{fig:extra_zeros}
\end{figure}

The estimated study-specific excess zero probabilities, $\pi_{ps}$, are shown in Figure~\ref{fig:extra_zeros}, along with 95\% credible intervals. These probabilities  ranged from $0-0.95$ and varied markedly across food items and studies.  Some food items (e.g., sugar, saccharin, other sweeteners) showed consistent zero inflation across studies, while others (e.g., mixed salad, yogurt, artichokes) exhibited different excess zero probabilities across studies with non-overlapping credible intervals, supporting the need for study-specific $\pi_{ps}$.


\begin{figure}
    \centering
    \includegraphics[width=0.8\linewidth]{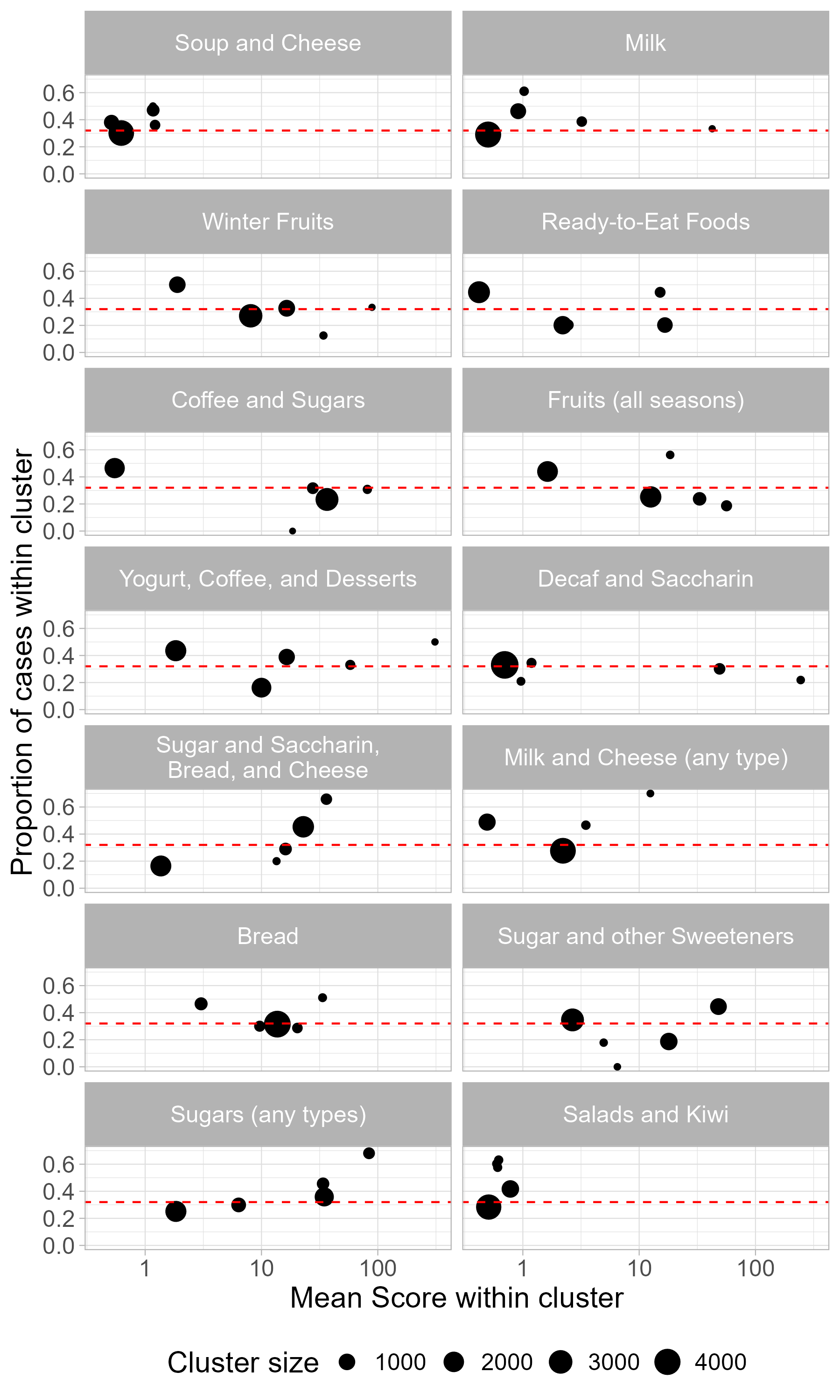}
    \caption{Exploratory associations between dietary patterns and cancer based on cluster assignments from the dependent Dirichlet process. For each cluster, we visualize mean pattern score, proportion of cases, and total cluster size.}
    \label{fig:exploratory_all}
\end{figure}


\FloatBarrier

\bibliographystyle{biom} \bibliography{biomsample}